\documentclass[12pt]{article}
\usepackage{amsfonts,amsmath,amssymb,amscd,mathrsfs}
\usepackage{breakcites}
\usepackage{graphicx}
\usepackage{textcomp}
\usepackage{float}

\usepackage[
backend=bibtex, 
natbib=true,
style=phys,
biblabel=brackets,
giveninits=true,
abbreviate=false,
isbn=false,
block=space,
]{biblatex}

\title{Absence of gravitationally induced entanglement in certain semi-classical theories of gravity }

\author{Ward Struyve\footnote{Department of Physics and Astronomy, KU Leuven, Belgium}$^{*}$\footnote{Centre for Logic and Philosophy of Science, KU Leuven, Belgium}  }

\def\lam{\lambda}

\def\pa{\partial}

\def\ii{\textrm i}
\def\ee{\textrm e}
\def\e{\textrm e}

\newcommand{\be}{\begin{equation}}
\newcommand{\en}{\end{equation}}
\newcommand{\bi}{\begin{itemize}}
\newcommand{\ei}{\end{itemize}}

\begin{document}
\date{}

\maketitle

\begin{abstract}
\noindent
Bose {\em et al.}\ and Marletto and Vedral proposed an experiment to test whether gravity can induce entanglement between massive systems, arguing that the capacity to do so would imply the quantum nature of gravity. In this work, a class of semi-classical models is examined that treat gravity classically, through some potential in the Schr\"odinger equation, and it is shown that these models do not generate entanglement. This class includes the Newton-Schr\"odinger model, where gravity is sourced by the wave function, the Bohmian analogue, where gravity is sourced by actual point-particles, and an interpolating model proposed by D\"oner and Gro{\ss}ardt. These models are analyzed in the context of the proposed experiment and contrasted with the standard Newtonian potential, which does generate entanglement.

\end{abstract}

\section{Introduction}
One of the central open problems in theoretical physics is the reconciliation of quantum theory with gravity. The quantum description of matter requires that the classical framework of general relativity must, at some level, be modified or extended. One possibility, which has been mostly investigated, is that gravity itself should be described by a quantum theory. This approach is pursued in for example the Wheeler-DeWitt theory, loop quantum gravity and string theory \cite{kiefer04}. An alternative possibility is that gravity remains fundamentally classical, in which case the question arises of how exactly the gravitational field couples to quantum matter. Various ideas for such semi-classical models have been explored, see e.g.\ \cite{hall16,tilloy16,tilloy18,struyve17a,donadi22,giulini23,oppenheim23}. 

To probe this problem experimentally, Bose {\em et al.}\ \cite{bose17} and Marletto and Vedral \cite{marletto17} (BMV) have proposed a feasible tabletop experiment to test whether gravity can induce entanglement between massive systems (see \cite{huggett23,bose25,marletto25} for recent reviews). They argued that the capacity to do so would imply the quantum nature of gravity. However, this claim has been challenged by several authors, who argued that the generation of entanglement does not necessarily constitutes evidence of quantumness \cite{hall18,anastopoulos20,doner22,carney22}. In particular, D\"oner and Gro{\ss}ardt~\cite{doner22} claimed to present a counterexample in the form of a model that treats gravity classically yet still generates entanglement. The aim of this paper is to show that their model in fact does not generate entanglement and hence does not constitute a counterexample.

More generally, in the context of the weak-field Newtonian regime, we will consider a class of semi-classical models where the gravitational interaction is modeled by a potential in the Schr\"odinger equation. The potential may depend on the wave function, leading to a non-linear dynamics. However, since the potential is additively separable, these models do not generate entanglement. The prime examples are theories where the potential is the Newtonian potential sourced by a particular mass density. This is the case in the Newton-Schr\"odinger (NS) model \cite{diosi84,bahrami14}, where gravity is sourced by the wave function, a Bohmian analogue \cite{struyve20a}, where gravity is sourced by the point-particles and a GRW-type analogue \cite{tilloy18}, where gravity is sourced by the flashes (i.e., the collapse events). The model proposed by D\"oner and Gro{\ss}ardt is an interpolation between the NS and NSB model and also involves a separable potential.

Since, the BMV experiment has not yet been realized, it is of interest to analyze and compare existing semi-classical models in this context. Here, we will examine the NS and NSB models and contrast them with the standard Newtonian potential which does generate entanglement, using the entanglement witness proposed by Chevalier {\em et al.} \cite{chevalier20}. These models predict a different behaviour of the witness, so that the BMV experiment could in principle discriminate between them.

\section{Absence of entanglement generation}\label{aeg}
In the models considered here, systems are described by a wave function $\psi_t(x)$, with $x=({\bf x}_1,\dots, {\bf x}_N)$, and a configuration $Q_t$, like for example particle positions. The wave function satisfies the non-relativistic Schr\"odinger equation
 \be
 \ii \hbar \pa_t \psi_t(x)  = \left[  -\sum^N_{i=1} \frac{\hbar^2  }{2m_i}\nabla_i^2 + V(x,t;\psi_t,Q_t)\right] \psi_t(x), \label{1}
 \en
with a potential that may depend on $\psi_t$ and $Q_t$, thereby allowing for a non-linear dynamics. The dynamics of $Q_t$ may also depend on $\psi_t$ and may be either deterministic or stochastic. 

In the standard quantum treatment, the potential is the Newtonian gravitational potential
\be
V_{\textrm N}(x) = - \frac{G}{2} \sum^N_{\substack{i,j=1 \\ i \neq j}} m_i m_j  \frac{1}{|{\bf x}_i - {\bf x}_{j}|}.
\en
In certain semi-classical theories, the potential is the Newtonian gravitational potential sourced by a particular mass density $\rho({\bf x},t)$, i.e.,
\be
V(x,t)= - G \sum_i m_i \int d^3 x \frac{\rho({\bf x},t) }{|{\bf x}_i - {\bf x}|}  , 
\label{19.001}
\en
Different choices for the mass density can be considered. In the NS model \cite{diosi84,bahrami14}, the mass density is 
\be
\rho_{\textrm{NS}} ({\bf x},t) = \sum^N_{i=1} m_i  \int  d^3 x'_1 \dots d^3 x'_N \delta({\bf x} - {\bf x}'_i)|{\psi}_t({\bf x}'_1, \dots,{\bf x}'_N )|^2,
\en
where $\psi$ is assumed to be normalized, resulting in the potential
\be
V_{\textrm{NS}}(x;{\psi}_t) = - G \sum^N_{i,j=1} m_i m_j \int  d^3 x'_1 \dots d^3 x'_N \frac{|{\psi}_t({\bf x}'_1, \dots,{\bf x}'_N )|^2}{|{\bf x}_i - {\bf x}'_j|}. \label{19}
\en
In the NSB model \cite{struyve20a}, the mass density is
\be
\rho_{\textrm{NSB}} ({\bf x},t) = \sum^N_{i=1} m_i \delta({\bf x} - {\bf X}_{i,t}), 
\en
where $X_t = ({\bf X}_{1,t}, \dots , {\bf X}_{N,t})$ is a configuration of particle positions at time $t$, with dynamics 
\be
\dot {\bf X}_{i,t} = \frac{\hbar}{m_i} {\textrm{Im}} \left( \frac{{\boldsymbol \nabla}_i \psi_t (x)}{\psi_t(x)}\right) \Bigg|_{x=X_t} . \label{guidance}
\en 
The corresponding potential is
\be
V_{\textrm{NSB}}(x;X_t) = - G \sum^N_{i,j=1} m_i m_j  \frac{1}{|{\bf x}_i - {\bf X}_{j,t}|}.
\label{20}
\en
In the GRW model with massive flashes \cite{tilloy18}, the density is 
\be
\rho({\bf x},t) = \sum^N_{i=1} m_i \sum_k \lam^{-1} f({\bf x} - {\bf x}_{i,k},t - t_{i,k} )
\en
where $({\bf x}_{i,k},t_{i,k})$ is the $k$-th flash for the $i$-th particle, $\lambda$ the collapse rate and $f$ a smearing function. Finally, in the model   
by D\"oner and Gro{\ss}ardt, the potential $V_R$ depends on both the wave function $\psi_t$ and a configuration $X_t$ satisfying \eqref{guidance}. It is of a form similar to \eqref{19.001} (but does not derive from a single mass density): 
\be
V_{R}(x;{\psi}_t,X_t) = - G \sum^N_{i,j=1} m_i m_j  \int d^3 x \frac{f_{ij}({\bf X}_{j,t} - {\bf x},\psi_t)}{|{\bf x}_i - {\bf x}|},
\en
where $f_{ij}$ are particular functions \cite[eq.\ (23)]{doner22} that lead to the NS or NSB model for particular choices. 

Whether or not the dynamics generates entanglement depends on the form of the potential. If the potential is additively separable, i.e., 
\be
V(x,t;\psi_t,Q_t) = \sum^N_{i=1} V_i({\bf x}_i,t;\psi_t,Q_t),
\en
then the dynamics does not generate entanglement. That is, an initially separable wave function
\be
\psi_0(x)= \prod^N_{i=1} \psi_{i,0}({\bf x}_i) \label{20.1}
\en
will remain separable at all times. This would be immediately clear if the potential did not depend on $\psi$, neither explicitly nor implicitly through a $\psi$-dependent dynamics of $Q$. But the statement is also valid in the case of a $\psi$-dependent potential. To see this, consider first a fixed $({\bar \psi}_t,{\bar Q}_t)$ (say for $t\geqslant 0$). Then, the Schr\"odinger equation
\be
\ii \hbar \pa_t \psi_t(x)  =\sum^N_{i=1} {\widehat H}_i\left(t;{\bar \psi}_t,{\bar Q}_t\right) \psi_t(x), \label{21}
\en
with
\be
{\widehat H}_i\left(t;{\bar \psi}_t,{\bar Q}_t\right)=-\frac{\hbar^2}{2m_i} \nabla_i^2 + V_i({\bf x}_i,t;{\bar \psi}_t,{\bar Q}_t),
\en
has a $\psi$-independent potential and does not generate entanglement. The solution for an initially separable wave function $\psi_0$ is given by 
\be
\psi_t (x) = \prod^N_{i=1} \psi_{i,t}({\bf x}_i), 
\en
where $\psi_{i,t}({\bf x})$ is the solution of
\be
\ii \hbar \pa_t \psi_{i,t}({\bf x})  =  {\widehat H}_i\left(t;{\bar \psi}_t,{\bar Q}_t\right) \psi_{i,t}({\bf x}), \label{21-2}
\en
with initial wave function $\psi_{i,0}$. Consider $({\bar \psi}_t,{\bar Q}_t)$ now to be a solution of \eqref{1} with $V$ an additively separable potential and with initial state $({\bar \psi}_0,{\bar Q}_0)$.  Then ${\bar \psi}_t$ will also be a solution of \eqref{21}. Since it was just established that \eqref{21} does not generate entanglement, ${\bar \psi}_t$ will not be entangled if ${\bar \psi}_0$ is not entangled, and hence \eqref{1} does not generate entanglement either.

Another way to establish this result is to consider the formal solution of \eqref{1} in terms of a Dyson series. This has been applied for particular semi-classical models in \cite{hall18,gruca24}. For an initial wave function $\psi_0$, the solution of \eqref{21} can be written as the Dyson series
\be
\psi_t(x) = U_t({\bar \psi},{\bar Q}) \psi_0(x) = {\mathcal T} \e^{-\ii \int^t_{t_0} dt'  \sum^N_{i=1} {\widehat H}_i\left(t';{\bar \psi}_{t'},{\bar Q}_{t'}\right)} \psi_0(x).
\label{1.1}
\en
Since 
\be
\left[{\widehat H}_i\left(t;{\bar \psi}_t,{\bar Q}_t\right),{\widehat H}_j\left(t';{\bar \psi}_{t'},{\bar Q}_{t'}\right)\right]=0,
\en
the unitary operator $U_t({\bar \psi},{\bar Q})$ can be written as 
\be
U_t({\bar \psi},{\bar Q})=\bigotimes^N_{i=1} U_{i,t}({\bar \psi},{\bar Q}),
\label{uni}
\en
with
\be
 U_{i,t}({\bar \psi},{\bar Q}) = {\mathcal T} \e^{-\ii \int^t_{t_0} dt' {\widehat H}_i\left(t';{\bar \psi}_{t'},{\bar Q}_{t'}\right)}.
\en
Because of \eqref{uni}, if $\psi_0$ is separable, then $\psi_t$ will be as well. Moreover, if $\psi_0$ is entangled, then the degree of entanglement, as measured by the entanglement entropy  \cite{bennett96}, will not change in time. As before, we can now consider $({\bar \psi}_t,{\bar Q}_t)$ to be a solution of \eqref{1} to establish that the dynamics \eqref{1} does not generate entanglement, nor changes the degree of entanglement in the case of an initially entangled state.

Among the potentials considered above, only the Newtonian potential is capable of generating entanglement. Potentials of the form \eqref{19.001}, like $V_{\textrm{NS}}$, $V_{\textrm{NSB}}$ and the one in the GRW-type model, are additively separable and do not generate entanglement. The same conclusion holds for the potential $V_R$ proposed by D\"oner and Gro{\ss}ardt.

While potentials like $V_{\textrm{NS}}$, $V_{\textrm{NSB}}$ and $V_R$ do not generate entanglement, it is not the case that systems evolve independently. In these cases, the dynamics of a single particle wave function will depend on the other particles. This dependence may be non-local and allow for faster-than-light signaling.

Finally, a word of caution regarding measurement statistics. In the NS model, measurement outcomes are determined by the wave function and the model needs to be supplemented by a collapse postulate or a spontaneous collapse mechanism so that the measurement statistics is governed by the usual Born rule \cite{bahrami14,derakhshani14}. In Bohmian semi-classical models, on the other hand, measurement outcomes are determined by the particle positions. These positions are generally {\em not} distributed according to the Born rule \cite[p.\ 10] {struyve20a}. This is in contrast to standard Bohmian mechanics. The reason is that in these semi-classical models there is back-reaction from the particles on the wave function, so that different initial positions lead to different wave function evolutions, which is absent in standard Bohmian mechanics. This implies that the statistical analysis is more involved, with a measurement statistics that is generally not described by the Born rule. That said, in the case of the BMV experiment---which we turn to in the next section---it does appear reasonable to use the Born rule.

\section{Experimental setup}
The BMV proposal to test gravitationally induced entanglement concerns two massive systems, separated by a distance $d$, which are each held in a spatial superposition of two states $|L\rangle$ and $|R\rangle$, separated by a distance $\delta$. The initial state has the product form 
\be
| \psi_0\rangle =  \frac{1}{2}\left( |L\rangle_1 + |R\rangle_1 \right) \left( |L\rangle_2 + |R\rangle_2 \right). 
\label{25} 
\en
The centers of the wave functions $\psi_{iL} ({\bf x}) =\langle {\bf x}|L\rangle_i $ and $\psi_{iR} ({\bf x}) =\langle {\bf x}|R\rangle_i $, $i=1,2$, are assumed to be aligned along the $x$-direction, represented by the unit-vector ${\bf e}_x$. Denoting these centers by ${\bf X}_{iL}$ and ${\bf X}_{iR}$, they are given by
\begin{align}
	{\bf X}_{1L} &= \left(- \frac{d}{2} - \frac{\delta}{2}\right) {\bf e}_x, \quad \: {\bf X}_{1R} = \left(- \frac{d}{2} + \frac{\delta}{2}\right) {\bf e}_x, \nonumber\\
	{\bf X}_{2L} &= \left( \frac{d}{2} - \frac{\delta}{2}\right) {\bf e}_x, \quad \quad {\bf X}_{2R} = \left( \frac{d}{2} + \frac{\delta}{2}\right) {\bf e}_x.
\end{align}
It is assumed that the wave functions $\psi_{iL} ({\bf x})$ and $\psi_{iR} ({\bf x})$, $i=1,2$, are well-localized so that their overlap is negligible and $| \psi_0\rangle$ is approximately normalized.  

Assuming that, during the run of the experiment, the kinetic part in the Hamiltonian can be ignored and that the spatial and temporal variation of the potential $V({\bf x}_1,{\bf x}_2,t)$ is negligible within the bulk of the support of each term $\psi_{1k}({\bf x}_1)\psi_{2l}({\bf x}_2)$ in \eqref{25}, $k,l=L,R$,  the dynamics will merely produce phase shifts for each term, given by 
\be
\ee^{- \ii \Delta_{kl} t /\hbar},
\en
with 
\be
\Delta_{kl} = V({\bf X}_{1k},{\bf X}_{2l}),
\en
so that 
\be
| \psi_t \rangle=  \frac{1}{2}\sum_{k,l = L,R} \ee^{- \ii \Delta_{kl} t /\hbar} |k\rangle_1 |l\rangle_2 .
\label{state}
\en
(In an actual experimental setup, there might be motion orthogonal to the $x$-direction, as in the case of a free-fall setup, but that could easily be taken into account.) In the case of an additively separable potential, we have
\be
\Delta_{kl} = V_1({\bf X}_{1k}) + V_2({\bf X}_{2l}),
\en
so that 
\be
| \psi_t \rangle = \frac{1}{2}\left[ \ee^{- \ii  V_1({\bf X}_{1L})t/\hbar } |L\rangle_1  + \ee^{- \ii  V_1({\bf X}_{1R})t/\hbar }|R\rangle_1 \right] \left[\ee^{- \ii  V_2({\bf X}_{2L})t/\hbar } |L\rangle_2  + \ee^{- \ii  V_2({\bf X}_{2R})t/\hbar }|R\rangle_2 \right],  \label{30} 
\en
which remains non-entangled.

In the case of the Newtonian potential, the quantum state is \cite{bose17}:
\be
|\psi_t\rangle = \frac{1}{2}\left[ \ee^{ \ii \frac{\gamma t}{d} } |L\rangle_1 |L\rangle_2+ 
\ee^{ \ii \frac{\gamma t}{d+\delta}} |L\rangle_1 |R\rangle_2  + \ee^{ \ii\frac{\gamma t}{d-\delta }} |R\rangle_1 |L\rangle_2  + \ee^{\ii \frac{\gamma t}{d }} |R\rangle_1 |R\rangle_2 \right],
\en
where $\gamma = G m_1 m_2/\hbar$. This state is entangled, except for particular values of $t$.

The NS and NSB model can easily be analyzed if the self-interaction is dropped from the potential, i.e., by restricting the sums over $i,j$ in \eqref{19} and \eqref{20} over $i\neq j$. In the two-particle case, this results in
\be
{\widetilde V}_{\textrm{NS}}({\bf x}_1,{\bf x}_2;{\psi}_t) = - G m_1 m_2 \int  d^3 x'_1  d^3 x'_2 |{\psi}_t({\bf x}'_1, {\bf x}'_2 )|^2\left(\frac{1}{|{\bf x}_1 - {\bf x}'_2|} + \frac{1}{|{\bf x}_2 - {\bf x}'_1|}\right) , \label{35}
\en
\be
{\widetilde V}_{\textrm{NSB}}({\bf x}_1,{\bf x}_2;{\bf X}_{1,t},{\bf X}_{2,t}) = - G  m_1 m_2  \left( \frac{1}{|{\bf x}_1 - {\bf X}_{2,t}|}+ \frac{1}{|{\bf x}_2 - {\bf X}_{1,t}| }\right).
\label{36}
\en

Consider first the NS model. Since the spatial variation of ${\widetilde V}_{\textrm{NS}}$ within the bulk of the support of each term $\psi_{1k}({\bf x}_1)\psi_{2l}({\bf x}_2)$, with $k,l=L,R$, can be assumed negligible, there will not be temporal variation either, at least over sufficiently short time intervals. This is because the wave functions $\psi_{1k}({\bf x}_1)\psi_{2l}({\bf x}_2)$ have approximately disjoint support, and hence if the dynamics merely generates relative phases, then $|\psi_t|^2 \approx |\psi_0|^2$. As a result, 
\be
|\psi_t\rangle  =  \frac{1}{2}\ee^{ \ii \frac{2\gamma t}{d} }\left(\ee^{ \ii \frac{\gamma t}{d+\delta}} |L\rangle_1 + \ee^{ \ii \frac{\gamma t}{d-\delta}}|R\rangle_1 \right) \left( \ee^{ \ii \frac{\gamma t}{d-\delta}}|L\rangle_2 +\ee^{ \ii \frac{\gamma t}{d+\delta}} |R\rangle_2 \right). 
\label{37} 
\en

In the case of the NSB model, the potential and hence the wave function dynamics depends on the actual trajectories ${\bf X}_{1,t}$ and ${\bf X}_{2,t}$. These will be in the support of one of the wave functions $\psi_{1k}\psi_{2l}$, say $\psi_{1m}\psi_{2n}$, and will remain so under mere relative phase development. Because of the narrowness of $\psi_{1m}\psi_{2n}$, the configuration can be approximated by its center, i.e., $({\bf X}_{1,t},{\bf X}_{2,t})\approx ({\bf X}_{1m},{\bf X}_{2n})$. There are four possible choices for the configuration $({\bf X}_{1 m},{\bf X}_{2n})$, each occurring with equal probability 1/4 over an ensemble (assuming that the initial configuration is distributed according to $|\psi|^2$). Each choice $({\bf X}_{1 m},{\bf X}_{2n})$ will entail a different phase development and hence a different wave function 
\be
| \psi^{mn}_t \rangle=  \frac{1}{2}\sum_{k,l = L,R} \ee^{- \ii \Delta^{mn}_{kl} t /\hbar} |k\rangle_1 |l\rangle_2
\label{200.001} ,
\en
where 
\be
\Delta^{mn}_{kl} = {\widetilde V}_{\textrm{NSB},1}({\bf X}_{1k};{\bf X}_{2n})+ {\widetilde V}_{\textrm{NSB},2}({\bf X}_{2l};{\bf X}_{1m})
\label{50},
\en
and
\be
{\widetilde V}_{\textrm{NSB},1}({{\bf x}_{1};{\bf X}_{2n}}) = - G  m_1 m_2   \frac{1}{|{\bf x}_1 - {\bf X}_{2n}|}, \quad {\widetilde V}_{\textrm{NSB},2}({{\bf x}_{2};{\bf X}_{1m}}) = - G  m_1 m_2   \frac{1}{|{\bf X}_{1m} - {\bf x}_2|}.
\en
This results in
\begin{align}
|\psi^{LL}_t\rangle  &=  \frac{1}{2}\left(\ee^{ \ii \frac{\gamma t}{d}} |L\rangle_1 + \ee^{ \ii \frac{\gamma t}{d-\delta}}|R\rangle_1 \right) \left( \ee^{ \ii \frac{\gamma t}{d}}|L\rangle_2 +\ee^{ \ii \frac{\gamma t}{d+\delta}} |R\rangle_2 \right), 	\nonumber\\
	|\psi^{LR}_t\rangle  &= \frac{1}{2}\left(\ee^{ \ii \frac{\gamma t}{d+\delta}} |L\rangle_1 + \ee^{ \ii \frac{\gamma t}{d}}|R\rangle_1 \right) \left( \ee^{ \ii \frac{\gamma t}{d}}|L\rangle_2 +\ee^{ \ii \frac{\gamma t}{d+\delta}} |R\rangle_2 \right), \nonumber\\
	|\psi^{RL}_t\rangle  &= \frac{1}{2}\left(\ee^{ \ii \frac{\gamma t}{d}} |L\rangle_1 + \ee^{ \ii \frac{\gamma t}{d-\delta}}|R\rangle_1 \right) \left( \ee^{ \ii \frac{\gamma t}{d-\delta}}|L\rangle_2 +\ee^{ \ii \frac{\gamma t}{d}} |R\rangle_2 \right), \nonumber\\
	|\psi^{RR}_t\rangle  &= \frac{1}{2}\left(\ee^{ \ii \frac{\gamma t}{d+\delta}} |L\rangle_1 + \ee^{ \ii \frac{\gamma t}{d}}|R\rangle_1 \right) \left( \ee^{ \ii \frac{\gamma t}{d-\delta}}|L\rangle_2 +\ee^{ \ii \frac{\gamma t}{d}} |R\rangle_2 \right). \label{50.01}
\end{align}
Hence, the ensemble is described by a statistical mixture of the (non-entangled) states $|\psi^{mn}\rangle$, each occurring with equal probability.{\footnote{The NSB model has also been investigated in \cite{andersen19}, where merely an approximate analysis was performed, which involved ignoring some of the phases. Unfortunately, this led to the incorrect conclusion that the state $|\psi^{RL}_t\rangle$ should be entangled.}} 

D\"oner and Gro{\ss}\-ardt also drop the self-interaction in the potential $V_R$ to analyze the BMV experiment. The potential is of the form:
\be
{\widetilde V}_{R}({\bf x}_1,{\bf x}_2;{\psi}_t,{\bf X}_{1,t},{\bf X}_{2,t}) = {\widetilde V}_{R,1}({\bf x}_1;{\psi}_t,{\bf X}_{2,t}) + {\widetilde V}_{R,2}({\bf x}_2;{\psi}_t,{\bf X}_{1,t})
\en
where
\be
{\widetilde V}_{R,1} = - G  m_1 m_2 
\int  d^3 x \frac{f_1 ({\bf x};{\psi}_t,{\bf X}_{2,t})}{|{\bf x}_1 - {\bf x}| }  ,\quad  V_{R,2} = - G  m_1 m_2 
\int  d^3 x \frac{f_2 ({\bf x};{\psi}_t,{\bf X}_{1,t})}{|{\bf x}_2 - {\bf x}| },
\label{50.5}
\en
with $f_1$ and $f_2$ particular functions \cite[eq.\ (28)]{doner22}. The potential is additively separable and hence it cannot generate entanglement. So why do D\"oner and Gro{\ss}\-ardt claim that it does? The claim appears to rely on an incorrect calculation of the wave function. Just as in the NSB model, the phases will depend on the actual trajectories, leading to four different wave functions. Denoting these wave functions again by $| \psi^{mn}_t \rangle$, where the labels $m,n$ refer to the actual trajectories $({\bf X}_{1,t},{\bf X}_{2,t})\approx ({\bf X}_{1m},{\bf X}_{2n})$, we have
\be
| \psi^{mn}_t \rangle=  \frac{1}{2}\sum_{k,l = L,R} \ee^{- \ii \Delta^{mn}_{kl} t /\hbar} |k\rangle_1 |l\rangle_2
\label{200} ,
\en
where now 
\be
\Delta^{mn}_{kl} = {\widetilde V}_{R,1}({\bf X}_{1k};{\psi}_0,{\bf X}_{2n}) + {\widetilde V}_{R,2}({\bf X}_{2l};{\psi}_0,{\bf X}_{1m})
\label{201.3}.
\en
These wave functions are separable. However, rather than considering these wave functions, D\"oner and Gro{\ss}ardt claim that the dynamics yields just a {\em single} wave function
\be
| \psi_t \rangle=  \frac{1}{2}\sum_{k,l = L,R} \ee^{- \ii \Delta^{kl}_{kl} t /\hbar} |k\rangle_1 |l\rangle_2, \label{201}
\en
which differs from the $|\psi^{mn}_t \rangle$ in \eqref{200}. The problem seems to be that instead of considering a fixed particle configuration $({\bf X}_{1 m},{\bf X}_{2n})$ to calculate the phase factor of each term $|k\rangle_1 |l\rangle_2$ in \eqref{201}, a different configuration---namely $({\bf X}_{1 k},{\bf X}_{2l})$---is used for each of these terms, which is incorrect.

In an actual experimental setup, the goal is to consider spin-1/2 particles and to accomplish the spatial separation between $|L \rangle$ and $|R \rangle$ via a Stern-Gerlach device. This means that there is a coupling between the spatial and spin degrees of freedom, implying the replacements $|L \rangle_i \to |L\rangle_i |\uparrow \rangle_i$ and $|R \rangle_i \to |R\rangle_i |\downarrow \rangle_i$, with $|\uparrow \rangle$ and $|\downarrow \rangle$ the spin-up and down states along the $z$-direction. The introduction of spin does not affect the relative phases just derived. In the case of the NS model, $\psi$ should be replaced by a two-particle spinor $\Psi$ satisfying the Pauli equation, and $|\psi|^2$ by $\Psi^\dagger \Psi$ in the potential \eqref{35}. In the NSB case, the particle dynamics~\eqref{guidance} should be replaced by 
\be
\dot {\bf X}_{i,t} = \frac{\hbar}{m_i} {\textrm{Im}} \left( \frac{\Psi^\dagger_t  (x){\boldsymbol \nabla}_i \Psi_t(x)}{\Psi^\dagger_t (x)\Psi_t (x)} \right) \Bigg|_{x=X_t}.
\en
But it is easy to see that these modifications do not affect the relative phases. After letting the phases develop for some time, the  states $|L \rangle_i$ and $|R \rangle_i$ should be recombined and appropriate spin measurements should be made to establish whether the state is entangled. This can be done with an entanglement witness. A suitable witness operator was proposed by Chevalier {\em et al.} \cite{chevalier20}:{\footnote{As pointed out in \cite{chevalier20}, the witness originally suggested in \cite{bose17} is not adequate for their proposed choice of parameters.}}
\be
{\widehat W} = I \otimes I - \sigma_x \otimes \sigma_x - \sigma_y \otimes \sigma_z - \sigma_z \otimes \sigma_y,
\en
with $\sigma_x,\sigma_y,\sigma_z$ the Pauli matrices. If $W = \langle \psi |{\widehat W}|\psi\rangle < 0 $, then $|\psi\rangle$ is entangled. (For separable states, $W \geqslant 0$. But also for entangled states it may be that $W \geqslant 0$.) For the state \eqref{state}, the witness is
\begin{multline}
W = 1 -\frac{1}{2} \big[ \cos(\gamma_{LL} - \gamma_{RR}) + \cos(\gamma_{LR} -\gamma_{RL} )\\ + \sin(\gamma_{LR} - \gamma_{LL}) + \sin(\gamma_{LR} - \gamma_{RR}) + \sin(\gamma_{RL} - \gamma_{LL}) + \sin(\gamma_{RL} - \gamma_{RR}) \big],
\end{multline}
where $\gamma_{kl}(t) = -\Delta_{kl}t/\hbar$. In the case of a separable potential, $\Delta_{kl} = \Delta_{1,k} + \Delta_{2,l}$, with $\Delta_{i,k}= V_i({\bf X}_{ik})$ and $\gamma_{i,k}(t) = -\Delta_{i,k}t/\hbar$, the witness becomes
\be
W = 1-\cos\left(\gamma_{1,L}-\gamma_{1,R}\right)\cos\left(\gamma_{2,L}-\gamma_{2,R}\right),
\en
which clearly satisfies $W \geqslant 0$.

\begin{figure}[t]
	\centering
	\includegraphics[width=0.6\textwidth]{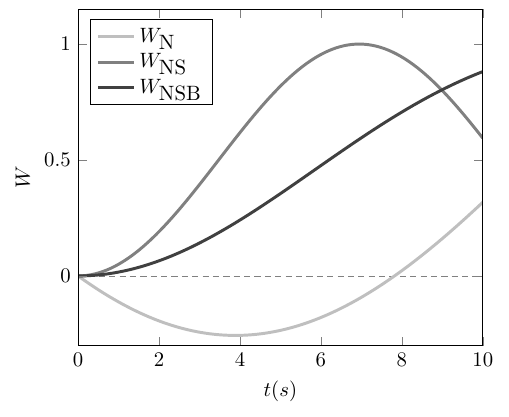}
	\caption{
		Entanglement witness $W$ evaluated for the different potentials $V_{\textrm{N}}$, ${\widetilde V}_{\textrm{NS}}$ and ${\widetilde V}_{\textrm{NSB}}$, plotted as a function of time. Negativity of the witness indicates entanglement. 
	}
	\label{fig:witness}
\end{figure}

In the case of the potential $V_{\textrm{N}}$, we have
\be
W_{\textrm{N}}(t) = \frac{1}{2}- \frac{1}{2} \cos\left(\frac{\gamma t}{d+\delta} - \frac{\gamma t}{d-\delta}\right) + \sin\left(\frac{\gamma t}{d} - \frac{\gamma t}{d+\delta}  \right)  + \sin\left( \frac{\gamma t}{d} - \frac{\gamma t}{d-\delta} \right) ,
\en
and in the case of ${\widetilde V}_{\textrm{NS}}$,
\be
W_{\textrm{NS}}(t) = 1-\cos^2\left( \frac{\gamma t}{d+\delta} - \frac{\gamma t}{d-\delta} \right).
\en
In the case of ${\widetilde V}_{\textrm{NSB}}$, the ensemble is a statistical mixture of the states $|\psi^{kl}\rangle$ given in \eqref{50.01} with equal probabilities. The corresponding witness $W_{\textrm{NSB}}$ is given by the average $\sum_{k,l = L,R}\langle \psi^{kl} |{\widehat W}|\psi^{kl}\rangle/4$:
\be
W_{\textrm{NSB}}(t) = 1- \frac{1}{4}\left[ \cos\left( \frac{\gamma t}{d} - \frac{\gamma t}{d+\delta} \right) +\cos\left( \frac{\gamma t}{d} - \frac{\gamma t}{d-\delta} \right)\right]^2.
\en
It is assumed that the measurement statistics is governed by the Born rule. While this assumption does not generally hold (as discussed at the end of section \ref{aeg}), it appears reasonable in the present context, because of the sharp localization of the wave packets, in particular at the final stage where the spin measurements take place.

The entanglement witnesses are plotted in fig.~\ref{fig:witness}, for the parameters originally considered in \cite{bose17}: $d=450\mu$m, $\delta=250\mu$m, $m_1=m_2=10^{-14}$kg. Decoherence effects have not been taken into account, but tend to become appreciable after $t\simeq 2$s \cite{bose17,chevalier20}. Detecting a negative witness (and hence entanglement) would favour the Newtonian potential, while a positive witness could be used to discriminate between the NS and NSB models.

\section{Conclusion}
We have shown that semi-classical theories for gravity with an additively separable potential do not generate entanglement. This class includes both the Newton–Schrödinger model and its Bohmian analogue. By contrast, the ordinary Newtonian potential is not additively separable and is capable of generating  entanglement. Applied to the BMV experiment, these models predict distinct entanglement witness signatures, allowing them to be experimentally distinguished.

\section{Acknowledgments}
It is a pleasure to thank Abdelrahman Azab, Thibaut Demaerel, Christian Maes, Xavier Oriols and Antoine Tilloy for helpful discussions and comments, and Simon Krekels for help with the figure. This work is supported by the Research Foundation Flanders (Fonds Wetenschappelijk Onderzoek, FWO), Grant No.\ G0C3322N. 

\printbibliography

\end{document}